\documentstyle[12pt]{article}

\newcommand{\be}[1]{\begin{equation} \label{(#1)}}
\newcommand{\ee}{\end{equation}}
\newcommand{\ba}[1]{\begin{eqnarray} \label{(#1)}}
\newcommand{\ea}{\end{eqnarray}}
\newcommand{\nn}{\nonumber}
\newcommand{\rf}[1]{(\ref{(#1)})}
\def\rp{$R_p \hspace{-1em}/\;\:$}

\def\rpt{$R_p \hspace{-0.85em}/\ \ $}

\def\pmb#1{\setbox0=\hbox{#1}%
  \kern-.015em\copy0\kern-\wd0
  \kern.03em\copy0\kern-\wd0
  \kern-.015em\raise.0233em\box0 }
\def\olrap{\overrightarrow{\partial}\hskip-4.6mm\overleftarrow{\partial}}
\def \znbb {0\nu\beta\beta}

\def\bfsgm{\pmb{${\sigma}$}}

\def\si{{ \bfsgm_{i}^{~}}}
\def\sj{{ \bfsgm_{j}^{~}} }
\begin{document}

\bigskip
\begin{center}
{\bf  R-parity Conserving Supersymmetry, Neutrino Mass
      and Neutrinoless Double Beta Decay.}
\bigskip

{M. Hirsch\footnotemark[1], H.V. Klapdor-Kleingrothaus\footnotemark[2]
\bigskip

{\it
Max-Plank-Institut f\"{u}r Kernphysik, P.O. 10 39 80, D-69029,
Heidelberg, Germany}

\bigskip
S.G. Kovalenko\footnotemark[3]
\bigskip

{\it Joint Institute for Nuclear Research, Dubna, Russia}
}
\end{center}

\begin{abstract}
We consider contributions of R-parity conserving softly broken
supersymmetry (SUSY) to neutrinoless double beta ($\znbb$) decay via
the (B-L)-violating sneutrino  mass term. The latter is a generic
ingredient of any weak-scale SUSY model with a Majorana neutrino mass.
The new R-parity conserving SUSY contributions to $\znbb$ are realized
at the level of box diagrams. We derive the effective Lagrangian describing
the SUSY-box mechanism of $\znbb$-decay and the corresponding nuclear
matrix elements. The 1-loop sneutrino contribution to the Majorana
neutrino mass is also derived.

Given the data on the $\znbb$-decay half-life of $^{76}$Ge and
the neutrino mass we obtain constraints on the (B-L)-violating sneutrino
mass. These constraints leave room for accelerator searches for certain
manifestations of the 2nd and 3rd generation (B-L)-violating sneutrino
mass term,
but are most probably too tight for first generation (B-L)-violating
sneutrino masses to be searched for directly.
\end{abstract}
\bigskip
\bigskip
\footnotetext[1]{M.Hirsch@mpi-hd.mpg.de}
\footnotetext[2]{KLAPDOR@ENULL.MPI-HD.MPG.DE}
\footnotetext[3]{KOVALEN@NUSUN.JINR.DUBNA.SU}

\newpage

\section{Introduction}

Neutrinoless double beta ($\znbb$) decay \cite{hax84,doi85} is a unique
example of a nuclear process which allows one to probe for lepton number
violation. Given the fact that the standard model conserves L at the
classical level, $\znbb$ decay can proceed only via non-SM interactions.
Therefore - taking into account the stringent experimental
limits \cite{hdmo97} - $\znbb$ decay is extremely sensitive to physics
beyond the standard model.

The simplest source of lepton number violation directly leading to $\znbb$
decay is a finite Majorana neutrino rest mass. It violates lepton number
by two units $\Delta L=2$ precisely what is necessary for $\znbb$-decay.
The corresponding contribution is described by the tree-level diagram of
the 4th order in the weak coupling constant with one Majorana neutrino
propagator as shown in Fig.1(a).

Supersymmetric extensions of the SM bring in new sources of lepton number
violation and, as a result, new mechanisms of $\znbb$-decay. Even with the
minimal matter and Higgs field content there are renormalizable L and B
violating terms in the superpotential and in the sector of soft-supersymmetry
breaking interactions which are not forbidden by gauge symmetry. These terms
violate not only L and B but also R-parity defined as $R_p = (-1)^{3B+L+2S}$
with $S$ being a particle spin. Models containing such interaction terms are
usually referred to as SUSY models with explicit $R_P$-breaking
\cite{R_P-explicit}. ($R_P$ can also be broken spontaneously via a non-zero
vacuum expectation value of the sneutrino field
$<\tilde\nu>\neq 0$ \cite{R_P-spon}.)

Formerly it was widely believed that supersymmetry can contribute to
$\znbb$ decay only if $R_P$ is broken \cite{MV}-\cite{HKK1}. The
corresponding mechanisms have been comprehensively studied in the last
few years \cite{HKK_top}-\cite{FKSS}.
In particular, it was  shown that the
current experimental limits from non-observation of $\znbb$-decay sets
upper bounds on certain \rpt Yukawa coupling constants which are more
stringent \cite{HKK_top,HKK} than previously known from
various accelerator and
non-accelerator experiments. Moreover, they turned out to be more stringent
than those from some forthcoming accelerator experiments. This conclusion
has put $\znbb$-decay forward as an interesting probe of supersymmetry.

Although there are no compelling theoretical arguments for R-parity
conservation, there exist a number of well-known phenomenological drawbacks
for supersymmetric models in which $R_P$ is violated. Maybe the most serious
one is the instability of the lightest SUSY particle. As a result, the
supersymmetric solution for the dark matter problem is lost unless
\rp Yukawa coupling constants $\lambda$ become unnaturally small, typically
$\lambda\leq 10^{-16}$.

In view of this and other problems for \rp SUSY there arises a natural
question whether $R_P$-violation is an inevitable condition for SUSY to
contribute to $\znbb$-decay. The present paper addresses this question.
We will demonstrate that there is a non-trivial R-parity conserving SUSY
contribution to $\znbb$-decay. 
In Fig.1(b) we present an example of a diagram associated
with the lowest 
order contribution to $\znbb$-decay within the R-parity conserving
minimal supersymmetric standard model (MSSM) with a Majorana neutrino
mass $m_M^{\nu}$. This particular example gives an explicit answer to
the above question: R-parity violation is not a necessary condition for
a SUSY contribution to $\znbb$-decay.

It is also obvious that this SUSY contribution is strongly suppressed
compared to the non-SUSY diagram in Fig.1(a). This is because it is of
higher order in perturbation theory, contains heavy sparticles in
intermediate states and receives a typical suppression due to the loop
integration. Moreover, this  diagram is proportional to the very small
factor $m_M^{\nu}/p_F$ where $p_F\approx 80$MeV is the nucleon Fermi 
momentum. The latter is also true for the simplest non-SUSY 
diagram in Fig. 1(a). The reason is common for both diagrams. In fact,
both the SM and the MSSM interactions conserve lepton number L. Therefore,
the only source for $\Delta L = 2$ violation, necessary for $\znbb$-decay
to proceed, is the Majorana neutrino mass term. If $m_M^{\nu} = 0$ lepton
number would be a conserved quantity and the $\znbb$-decay amplitude
should vanish, $R_{\znbb} = 0$. Such a behavior corresponds to
$R_{\znbb} \sim m_M^{\nu}/p_F$ in the limit of small $m_M^{\nu}$.

Thus, the diagram in Fig.1(b), given its very small contribution to
$\znbb$-decay, provides just a principal demonstration of the fact
that $\znbb$-decay can be triggered by R-parity conserving supersymmetry.
No practical consequences can be obtained from this new
diagram in the sense of establishing new constraints either on SUSY
parameters or on $m_M^{\nu}$ from non-observation of $\znbb$-decay.

However, we will show that there are other R-parity conserving SUSY
contributions via the lepton number violating sneutrino mass term.
As was shown in \cite{theorem} the Majorana neutrino mass, the
(B-L)-violating sneutrino mass and the $\znbb$-decay amplitude are
generically connected to each other. Namely, non-vanishing of one of
these three quantities implies non-zero values of the remaining two.
Thus, the $\znbb$-amplitude should always contain a contribution
corresponding to the (B-L)-violating sneutrino mass term.

We will study this contribution and extract constraints on the
(B-L)-violating sneutrino mass term from the current experimental data
on $\znbb$-decay of $^{76}$Ge.

This paper is organized as follows. In section 2 we give a short account
on the structure of neutrino-sneutrino mass terms and the theorem which
establishes the above mentioned relations between the neutrino and
sneutrino masses and $\znbb$-decay. Sect.3 is devoted to some general
properties of possible SUSY contributions to $\znbb$-decay. In this
section we specify the box diagrams describing the $R_P$-conserving SUSY
contribution. Our approach to the derivation of the corresponding
$\znbb$-transition operators and nuclear matrix elements is outlined in
sect. 4. Sect. 5 deals with $\znbb$-decay  constraints on the
(B-L)-violating sneutrino mass. In section 6 we calculate the sneutrino
contribution to the Majorana neutrino mass and derive then limits from
experimental data on neutrino masses. We then close with a short
summary.

\section{Structure of the Neutrino-Sneutrino Mass Terms}

As shown in \cite{theorem}, the self-consistent form of the neutrino
and sneutrino mass terms is
\ba{complete}
{\cal L}^{\nu\tilde\nu}_{mass} = -\frac{1}{2} (m_M^{\nu}
\overline{\nu^c_L} \nu_L + h.c.)  - \frac{1}{2}(\tilde m_M^2
\tilde\nu_L \tilde\nu_L + h.c.) - \tilde m_D^2 \tilde\nu_L^* \tilde\nu_L.
\ea
where $\nu = \nu^c$ is a Majorana field. The first two terms violate
the global (B-L) symmetry while the last one respects it. The first
term is a Majorana mass term of the neutrino. We  call the second term
a "Majorana"-like mass, while the third one is referred to as a
"Dirac"-like sneutrino mass term. This reflects an analogy with Majorana
and Dirac mass terms for neutrinos. In the presence of the right handed 
neutrino field $\nu_R$ the Dirac neutrino mass term 
$m_D^{\nu} (\bar\nu_L \nu_R + \bar\nu_R\nu_L)$ could also be included
in Eq. \rf{complete}  but it is  not required by the self-consistency
arguments. Note that $\tilde m_M^2$ is not a positively defined
parameter.

Eq. \rf{complete} is a generic consequence of weak-scale softly broken
supersymmetry and does not depend on the specific mechanism of mass
generation in the low-energy theory. For the sake of simplicity and
without any loss of generality we ignore possible neutrino mixing.

The low-energy theorem proven in ref. \cite{theorem} relates the
following three (B-L)-violating quantities: the neutrino Majorana
mass $m_M^{\nu}$, the "Majorana"-like sneutrino mass $\tilde m_M$
and the amplitude of $0\nu\beta\beta$-decay $R_{0\nu\beta\beta}$.
Here we shortly describe the proof of this theorem.

It is relatively easy to see that if at least one of the quantities
is non-zero the two others are generated in higher orders of perturbation
theory as demonstrated in Fig. 2, where only dominant diagrams are
shown. Internal lines in these diagrams are neutralinos $\chi_i$,
gluinos $\tilde g$, charginos $\chi^{\pm}$, selectron $\tilde e$,
u-squark $\tilde u$ and sneutrino $\tilde\nu$.
The latter is to be identified with the (B-L)-violating
``Majorana''  propagator proportional to $\tilde m_M^2$.
The sneutrino ``Majorana'' propagator was explicitly
derived in ref. \cite{theorem} and will be given below.

The various diagrams lead to relations among the three (B-L)-violating
observables, which we write down schematically
\ba{1-loop}
z_i &=& \sum_{i\neq j} a_{ij}\cdot z_j +  {\cal A}_i.
\ea
Here, $z_i$ can stand for $z_i = m_M^{\nu}$, $\tilde m_M^2$,
$R_{0\nu\beta\beta}$. The coefficients $a_{ij}$ correspond to
contributions of the diagrams in Fig.2(a)-(f) so that $i,j = a,b,c,d,e,f$.
Terms ${\cal A}_i$ represent any other possible contributions. The explicit
form of $a_{ij}$ and ${\cal A}_i$ is not essential in the following.
Important is only the presence of a correlation between
$m_M^{\nu}, \ \tilde m_M^2, \ R_{0\nu\beta\beta}$, expressed by
Eq. \rf{1-loop}.

Now we are going to prove that if $z_{i_1} = 0$ then  $z_{i_2} =
z_{i_3} = 0$ (the same will be true for any permutation). On the
basis of Fig. 2 and Eqs. \rf{1-loop} one can expect such properties
of the set of observables $z_i$.  Indeed $z_{i_1} = 0$ in the
left-hand side of Eq. \rf{1-loop} strongly disfavors $z_{j_2}\neq 0$
and $z_{j_3}\neq 0$, because it requires either all the three terms
in the right-hand sides to vanish or their net cancelation.
The latter is "unnatural". Even if such a cancelation would be done by
hand, using (unnatural) fine-tuning
of certain parameters, in some specific order of perturbation theory,
it would be spoiled again in higher orders of perturbation theory.
The cancelation of all terms in the right-hand side of Eqs. \rf{1-loop}
in all orders of perturbation theory  could only be
guaranteed by a special unbroken symmetry.
Let us envisage this possibility in details.

The effective Lagrangian of a generic model of weak scale softly broken
supersymmetry contains after electro-weak symmetry breaking the following
terms \cite{Haber}
\ba{L1}
{\cal L} &=& - \sqrt{2} g \epsilon_i \cdot
\overline{\nu}_L\chi_i\tilde\nu_L
- g \epsilon_i^- \cdot \overline{e}_L\chi_i^-\tilde\nu_L
- g \epsilon_i^+ \cdot \overline{\nu}_L\chi_i^+\tilde e_L + \\ \nonumber
&+&  \frac{g}{\sqrt{2}} (\overline{\nu}_L \gamma^{\mu} e_L +
\overline{u}_L \gamma^{\mu} d_L)W^+_{\mu}
+ g \cdot \bar\chi_i\gamma^{\mu} (O^L_{ij}P_L +
                                  O^R_{ij}P_R)\chi^+_j W^-_{\mu} \\ \nn
& + & ... + h.c.
\ea
Dots denote other terms which are not essential for our further
considerations. Here, $\tilde\nu_L$ and $\tilde e_L$ represent scalar
superpartners of the left-handed neutrino $\nu_L$ and electron $e_L$
fields. The chargino $\chi^{\pm}_i$ and neutralino $\chi_i$ are
superpositions of the gaugino and the higgsino fields. The contents
of these superpositions depend on the model. Note that the neutralino
is a Majorana field $\chi_i^c = \chi_i$. The explicit form of the
coefficients $\epsilon_i, \ \epsilon^{\pm}_i $ and $O^{L,R}_{ij} $ is
also unessential. For the case of the MSSM one can  find them, for
instance in \cite{Haber}. Eq. \rf{L1} is a general consequence of the
underlying weak scale softly broken supersymmetry and the spontaneously
broken electro-weak gauge symmetry.

The Lagrangian \rf{L1} does not posses any continuous symmetry having
non-trivial (B-L) transformation properties. Recall, that $U(1)_{B-L}$
is assumed to be broken since we admit (B-L)-violating mass terms in
Eq. \rf{complete}. However, there might be an appropriate unbroken
discrete symmetry. Let us specify this discrete symmetry
group by the following field transformations
\ba{discrete}
\nu &\rightarrow & \eta_{\nu} \nu, \ \ \
\tilde\nu\rightarrow\eta_{\tilde\nu} \tilde\nu, \ \ \
e_L \rightarrow  \eta_{e} e_L, \ \ \
\tilde e_L  \rightarrow \eta_{\tilde e} \tilde e_L, \\  \nonumber
W^+ & \rightarrow & \eta_{_W} W^+,  \ \ \
\chi_{i}\rightarrow\eta_{\chi_{i}} \chi_{i},\ \ \
\chi^+  \rightarrow  \eta_{\chi^+} \chi^+ , \ \ \
q_L \rightarrow \eta_{q} q_L .
\ea
Here $\eta_i$ are phase factors.
Since the Lagrangian \rf{L1} is assumed to be invariant under these
transformations one obtains the following relations
\ba{symm_rel}
\eta_{\nu}^* \eta_{\tilde\nu} \eta_{\chi_i} &=& 1, \ \ \
\eta_{e} \eta_{\chi^+}\eta_{\tilde\nu}^* = 1, \ \ \ ...\\ \nonumber
\eta_{e} \eta_{_W} \eta_{\nu}^* &=& 1, \ \ \
\eta_W^* \eta_{\chi^+} \eta_{\chi_i}^* = 1, \ \ \
\eta_d \eta_W \eta_u^* =1 \ \ \ ....
\ea
Dots denote other relations which are not essential here.
The complete set of these equations defines the admissible
discrete symmetry group of the Lagrangian in Eq. \rf{L1}.

Let us find the transformation property of the operator structure
responsible for $0\nu\beta\beta$-decay under this group.
At the quark level $0\nu\beta\beta$-decay implies the transition
$dd\rightarrow uuee$,  described by the effective operator
\ba{operator}
{\cal O}_{0\nu\beta\beta} = \alpha_i\cdot \bar u
\Gamma_i^{(1)} d\cdot \bar u \Gamma_i^{(1)} d\cdot
\bar e \Gamma_i^{(2)} e^c,
\ea
where $\alpha_i$ are numerical constants,  $\Gamma_i^{(k)}$ are
certain combinations of Dirac $\gamma$ matrices. The
$0\nu\beta\beta$-decay amplitude $R_{0\nu\beta\beta}$ is related to
the matrix element of this operator
\begin{eqnarray} \label{ampl}
R_{0\nu\beta\beta} \sim <2 e^- (A, Z+2)| {\cal O}_{0\nu\beta\beta} |(A,Z)>
\end{eqnarray}
where $(A,Z)$ is a nucleus with the atomic weight $A$ and the total
charge $Z$. The operator in Eq. \rf{operator} transforms under
the group \rf{discrete} as follows
\ba{operator_trans}
{\cal O}_{0\nu\beta\beta}\rightarrow
\eta^2_{0\nu\beta\beta}{\cal O}_{0\nu\beta\beta}
\ea
 with
\ba{eta_dbd}
\eta_{0\nu\beta\beta} =  \eta_d\eta_u^*\eta_e^*
\ea

 Solving Eqs. \rf{symm_rel}, \rf{eta_dbd}, one finds
\ba{sol}
\eta_{\nu}^2 = \eta_{\tilde\nu}^2 = (\eta^*_{0\nu\beta\beta})^2.
\ea
This relation proves the statements 1,2.  To see this we note
that the observable quantity
$z_i = (m_M^{\nu}, \ \tilde m_M^2, \ R_{0\nu\beta\beta})$
is forbidden by this symmetry if the corresponding discrete group
factor is non-trivial, {\it i.e.}  $\eta_i^2\neq 1$. Contrary, if
$\eta_i^2= 1$, this quantity is not protected by the symmetry
and appears in higher orders of perturbation theory, even if
it is not included at the tree-level.  Relation \rf{sol}
claims that if one of the $z_i$ is forbidden then the two others
are  also forbidden and, vice versa, if one of them is not forbidden
they are all not forbidden.

This completes the proof of the theorem relating the neutrino Majorana
mass $m_M^{\nu}$, the "Majorana"-like sneutrino mass $\tilde m_M$
and the amplitude of $0\nu\beta\beta$-decay $R_{0\nu\beta\beta}$.
The proven theorem can be considered as a supersymmetric
generalization of the well-known theorem \cite{SV} relating only
neutrino Majorana mass and the neutrinoless double beta decay
amplitude.

Let us show that the "Dirac"-like (B-L)  conserving sneutrino mass
$\tilde m_D$ should also be present in the theory to ensure
the stability of the vacuum state. Towards this end consider the last
two terms of Eq. \rf{complete} which we denote as
${\cal L}^{\tilde\nu}_{mass}$ and use the real field representation
for the  complex scalar sneutrino field
\ba{re_im}
\tilde\nu = (\tilde\nu_1 + i \tilde\nu_2)/\sqrt{2},
\ea
where $\tilde\nu_{1,2}$ are real fields. Then
\ba{snu_DM}
{\cal L}^{\tilde\nu}_{mass} =  - \frac{1}{2}(\tilde m_M^2
\tilde\nu_L \tilde\nu_L + h.c.) - \tilde m_D^2 \tilde\nu_L^*
\tilde\nu_L   = - \frac{1}{2}  \tilde m_1^2
\tilde\nu_1^2 -  \frac{1}{2}  \tilde m_2^2 \tilde\nu_2^2
\ea
where
\ba{split}
\tilde m_{1,2}^2 = \tilde m_D^2 \pm |\tilde m_M^2|
\ea
Assume the vacuum state is stable. Then $\tilde m_{1,2}^2\geq 0$, i.e.
$\tilde m_D^2 \geq |\tilde m_M^2|$, otherwise the vacuum is unstable and
subsequent spontaneous symmetry breaking occurs via non-zero
vacuum expectation values of the sneutrino fields $<\tilde\nu_i>\neq 0$.
The broken symmetry in this case is the R-parity. It is a discrete
symmetry defined as $R_p = (-1)^{3B+L+2S}$, where $S,\ B$ and $L$
are the spin, the baryon and the lepton quantum number.

Therefore, as indicated at the beginning of this section the self-consistent
structure of the mass terms of the neutrino-sneutrino sector is given by
Eq. \rf{complete}. The mass parameter $\tilde m_M$ gives a measure of
sneutrino-antisneutrino mixing ($\tilde\nu-\tilde\nu^*$). This
(B-L)-violating effect is an evident manifestation of the 2nd term
in Eq. \rf{complete}. On the other hand, a
finite $\tilde m_M$ gives rise to splitting the complex scalar field
${\tilde \nu} = ({\tilde \nu}_1 + i {\tilde \nu}_2)/\sqrt{2}$
into two real mass eigenfields ${\tilde \nu}_{1,2}$
with the masses ${\tilde m}_{1,2}^2 = {\tilde m}_D^2 \pm |{\tilde m}_M^2|$.
According to the above definition, ${\tilde \nu}_{1}$ is the CP-even
state while ${\tilde \nu}_{2}$ is the CP-odd one.

Let us write down the explicit form of the above mentioned
(B-L)-violating ``Majorana'' propagator $\Delta_{\tilde\nu}^M$ for
the sneutrino \cite{theorem} which is necessary for our subsequent
considerations.  It can be derived by the use of the real field
representation as in Eq. \rf{snu_DM}. For comparison we also give
the (B-L)-conserving ``Dirac'' $\Delta_{\tilde\nu}^D $  sneutrino
propagator,
\ba{propagators_1}
\Delta_{\tilde\nu}^D(x-y) =
i <0|T(\tilde\nu_L(x) \tilde\nu_L^{\dagger}(y)|0> =
\frac{1}{2}(\Delta_{\tilde m_1}(x-y) + \Delta_{\tilde m_2 }(x-y)),  \\
\Delta_{\tilde\nu}^M(x-y)=
i <0|T(\tilde\nu_L(x) \tilde\nu_L(y)|0> =
= \frac{1}{2}(\Delta_{\tilde m_1}(x-y) - \Delta_{\tilde m_2 }(x-y)),
\ea
where
\ba{def_1}
\Delta_{\tilde m_i}(x) = \int\frac{d^4 k}{(2\pi)^4}
\frac{e^{-ikx}}{\tilde m_i^2 - k^2 - i\epsilon}
\ea
is the ordinary propagator for a scalar particle with mass
$\tilde m_i$. Using the definition of $\tilde m_{1,2}$ as in
Eq. \rf{snu_DM} one finds
\ba{propagators_2}
\Delta_{\tilde\nu}^D(x) &=& \int\frac{d^4 k}{(2\pi)^4}
\frac{\tilde m_D^2 - k^2}{(\tilde m_1^2 - k^2 - i\epsilon)
(\tilde m_2^2 - k^2 - i\epsilon)} e^{-ikx}, \\
\label{(propagator_M)}
\Delta_{\tilde\nu}^M(x) &=& -  \tilde m_M^2 \int\frac{d^4 k}{(2\pi)^4}
\frac{e^{-ikx}}{(\tilde m_1^2 - k^2 - i\epsilon)
(\tilde m_2^2 - k^2 - i\epsilon)}.
\ea
It is seen that in absence of the (B-L)-violating sneutrino
``Majorana''-like mass term $\tilde m_M^2 =0$ the (B-L)-violating
propagator $\Delta_{\tilde\nu}^M$  vanishes while the (B-L)-conserving
one $\Delta_{\tilde\nu}^D$  becomes the ordinary propagator of a scalar
particle with mass $\tilde m_{1} = \tilde m_{2} = \tilde m_D$.

Majorana neutrino fields can propagate in a virtual state
conserving the (B-L) quantum number as well as  violating it.
In the Majorana representation of the neutrino field
\ba{majorana}
\nu = P_L \nu_L + P_R (\nu_L)^c
\ea
where $\nu^c = \nu$, the corresponding (B-L)-conserving  ($S^D$) and
(B-L)-violating ($S^M$) propagators can be written as
\ba{nu_prop}
S^D(x-y) &=& i <0|T(\nu_L(x) \bar\nu_L(y)|0> =
i P_L <0|T(\nu(x) \bar\nu(y)|0>P_R = \\ \nn
&=&  \int \frac{d^4k}{(2\pi)^4}
\frac{\gamma_{\mu} k^{\mu}}{m_{\nu}^2 - k^2 - i\epsilon}
e^{-ik(x-y)},\\
S^M(x-y) &=& i <0|T(\nu_L(x) \overline{\nu_L^c}(y)|0> =
P_L <0|T(\nu(x) \bar\nu(y)|0>P_L = \\ \nn
&=&  m_{\nu} \int \frac{d^4k}{(2\pi)^4}
\frac{1}{m_{\nu}^2 - k^2 - i\epsilon}
e^{-ik(x-y)},
\ea
where $m_{\nu} \equiv m_M^{\nu}$ for simplicity.
Therefore, the effect of (B-L)-violation originating from the
neutrino propagator is proportional to the Majorana neutrino mass
while a (B-L)-conserving contribution to $\znbb$ decay via neutrino
propagation in the Dirac mode does essentially not depend on the neutrino
mass and leads to a contribution proportional to the mean neutrino momentum
in a nucleus. The latter is typically of the order of the Fermi momentum
$\sim p_F \approx 80$MeV. As a result such a contribution, if it exists,
is greatly enhanced compared to the Majorana mass contribution.

\section{MSSM contribution to $\znbb$-decay. General properties and
                                             Effective Lagrangian}

In this section we are considering the general properties of the
$R_P$-conserving MSSM contribution to $\znbb$-decay and derive
the corresponding effective Lagrangian in terms of color-singlet
quark charged currents.

Within our supersymmetric framework there are two sources of lepton
number violation, the Majorana neutrino mass, $m_M^{\nu}$, and the
(square of the) ``Majorana''-like sneutrino mass, ${\tilde m}_M^2$.
Both of them violate - by construction - lepton number by two units.
$0\nu\beta\beta$ decay also violates $L$ by two units. The
$0\nu\beta\beta$ amplitude $R_{0\nu\beta\beta}$ will therefore be
proportional
\be{r0np}
R_{0\nu\beta\beta} \sim m_M^{\nu} {\cal O}^{(1)}  +
                      {\tilde m}_M^2 {\cal O}^{(2)},
\ee
with ${\cal O}^{(i)}$ representing some matrix elements. Contributions
proportional to $m_M^{\nu}$ can be classified by
${\cal O}^{(1)} = {\cal O}^{(1)}_{(SM)} + {\cal O}^{(1)}_{(SUSY)}$,
where the first part stands symbolically for the usual mass mechanism
diagram, see Fig. 1(a), while the second part summarizes all kinds of
diagrams involving virtual SUSY particles. An example of this type of
contribution was given in the introduction, see Fig. 1(b). Clearly,
all diagrams of ${0\nu\beta\beta}$ decay involving SUSY particles must
have at least 6 basic vertices. They are thus of higher order compared
to Fig. 1(a) and can be safely neglected:
${\cal O}^{(1)} = {\cal O}^{(1)}_{(SM)} + {\cal O}^{(1)}_{(SUSY)} \cong
{\cal O}^{(1)}_{(SM)}$.

Let us consider $R_{0\nu\beta\beta}$ in more details. At the quark level
neutrinoless double beta decay is induced by the transition of two $d$
quarks into two $u$ quarks and two electrons. This process is schematically
represented by the diagram in Fig. 3(a) encoding all possible contributions
to $\znbb$-decay. In the $R_P$ conserving supersymmetric model it is
useful to decompose the basic diagram Fig. 3(a) as shown in Fig. 3(b-g).

Given that the low-energy theory contains a light neutrino there are
always contributions involving a long-distance interaction component
associated with the light neutrino exchange. Let us use this fact for
a decomposition of the quark-lepton $\znbb$-effective vertex in Fig.
3(a) into different parts explicitly showing the presence of the neutrino
line, Fig. 3(b-d). It is also instructive to isolate the SUSY contributions
in the form of effective vertices induced by heavy SUSY particles
exchange. This decomposition is indicated by the small black spots
corresponding to the short-distance, approximately point-like, effective
SUSY induced interactions, Fig. 3(c-g). The first diagram, Fig. 3(b),
corresponds to the conventional standard model Majorana neutrino exchange
contribution mentioned above. The crossed neutrino line indicates the
lepton number violating Majorana neutrino propagator, $S^M$.
Diagrams Fig. 3(c,d) correspond to the neutrino accompanied SUSY
contributions. Here, neutrino lines are uncrossed and correspond to
lepton number conserving neutrino propagators, $S^D$. (As noted above
we neglect SUSY diagrams with lepton number violating propagators
proportional to the small neutrino mass $m_M^{\nu}$.) The last three
diagrams represent the purely supersymmetric contribution and are of
short-ranged nature.

This decomposition allows one to apply, if necessary, a Fierz
rearrangement to the approximately point-like black-spot SUSY vertices
in Fig. 3 and represent them in the form of a product of color-singlet
quark charged currents and a leptonic part. Such a representation is
crucial for the derivation of the $\znbb$-transition operators in the
non-relativistic impulse approximation and for the subsequent nuclear
structure calculations discussed in the sect. 4.

Assuming the Fierz rearrangement applied to the point-like SUSY vertices
one can write down the general form of the effective Lagrangian
reproducing the decomposition in Fig. 3 within 4th order of perturbation
theory. It can be written in the form:
\ba{effective}
{\cal L}_{\znbb} &=&  {\cal L}_{W{\bar f}f} +
\frac{\lambda^{(1)}_{i}}{m_{SUSY}^{n_1}} j_i \cdot
{\bar e \Gamma_{i(1)} \nu_L^c} +
+ \frac{\lambda_{i}^{(2)}}{m_{SUSY}^{n_2}} W^-_{\mu}\cdot \bar e
\Gamma_{i(2)}^{\mu} \nu_L^c
 \\  \nn
& + &
 \frac{\lambda^{(3)}_{i}}{m_{SUSY}^{n_3}}W^-_{\mu} W^-_{\nu}\cdot
\bar e \Gamma_{i(3)}^{\mu\nu} e^c
+
 \frac{\lambda^{(4)}_{i}}{m_{SUSY}^{n_4}}j_i^{\mu} W^-_{\mu} \cdot
\bar e \Gamma_{i(4)} e^c
 \\  \nn
& + &\frac{\lambda^{(5)}_{ij}}{m_{SUSY}^{n_5}} j_i j_j\cdot \bar e
\Gamma_{ij(5)} e^c,
\ea

where the SM term ${\cal L}_{W{\bar f}f}$ is also introduced (see Eq.
\rf{lcc} in Appendix A). Color-singlet local diquark operators are
defined as
\ba{q_cur}
j_i = \bar u^{\alpha} {\cal O}_i d_{\alpha}
\ea
with $\alpha$ being a color index. The objects $\Gamma _{i(k)}$ and
${\cal O}_i$ are constructed of Dirac gamma matrices as well as
derivatives.

Effective couplings $\lambda^{(k)}$ are dimensionless constants.
Different terms are scaled out by the characteristic SUSY breaking
mass scale $m_{SUSY}$ with an appropriate degree $n_i$ to
accommodate correct physical dimension of the corresponding term.
As seen from the leptonic part of the effective Lagrangian \rf{effective},
the first term conserves lepton number ($\Delta L = 0$) while the
remaining five terms violate it by two units ($\Delta L =2$).

In the effective Lagrangian \rf{effective} we neglected possible
L-conserving terms with the lepton operator structure
$\sim \bar e \Gamma_{i} \nu_L$.  Their contributions
to the $\znbb$-amplitude are strongly suppressed compared to
the contributions of the similar L-violating terms
$\sim \bar e \Gamma_{i} \nu_L^c$.
To see this fact let us have a closer look at
the corresponding leading order diagrams in Fig. 3(c,d).
The bottom parts of these
diagrams are the SM charged current (SMCC) interactions of the form
$
(\bar u \gamma_{\mu} P_L d)(\bar e \gamma^{\mu} P_L \nu_n)\cdot U_{en}
$, 
while the top parts correspond to the effective SUSY vertices.
If they are given by the 2nd and 3rd terms of
the effective Lagrangian Eq. \rf{effective}, the resulting leptonic
tensor ${\cal L}_{SUSY}$ can then be written
schematically as
\ba{lep}
{\cal L}_{\Delta L = 2} \sim \bar e \gamma_{\mu} P_L <0|T(\nu_k \bar \nu_n)|0>
P_R \Gamma e^c \cdot U_{ek}U^*_{en} \sim
\bar e \gamma_{\mu}\gamma_{\nu} P_R \Gamma e^c\cdot q^{\nu}/q^2,
\ea
where $q$ is the neutrino momentum. In the right hand side neutrino
masses $m_M^{\nu_n}$ are neglected since $m_M^{\nu_n} << \langle q \rangle $,
where $\langle q \rangle\sim p_F\approx 100$MeV is the average momentum
of a neutrino propagating in a nucleus ($p_F$ is the nucleon Fermi
momentum). The mixing matrix elements disappear on the right hand side
due to the unitarity relation $U_{ek}U^*_{en}\delta_{kn} = 1$.
This should be compared with the contribution of the possible L-conserving
terms $\sim \bar e \Gamma_{i} \nu_L$ which we neglected
in the effective Lagrangian in \rf{effective}. The leptonic tensor
in this case takes the form
\ba{mass_mech}
{\cal L}_{\Delta L = 0} \sim \bar e \Gamma_{i}
P_L <0|T(\nu_k \bar \nu_n)|0>
P_L \gamma_{\rho} e^c \cdot U_{ek}U_{en} \sim
\bar e \Gamma_{i}\gamma_{\rho} P_R e^c\cdot \langle m_{\nu} \rangle/q^2,
\ea
where $\langle m_{\nu}\rangle = m_M^{\nu_n} U_{en}^2$.
The same structure appears in the standard neutrino mass mechanism
with the SMCC at both ends of the virtual neutrino line. 

Comparing Eqs. \rf{lep} and \rf{mass_mech} one can see that
the SUSY contribution corresponding to the $\Delta L = 2$ operators
receives from the leptonic sector a huge enhancement compared to
the contribution of the $\Delta L = 0$ operators. In fact
\ba{enh}
 {\cal L}_{\Delta L = 2}/{\cal L}_{\Delta L = 0} \sim p_F/
\langle m_{\nu}\rangle\sim 10^8\cdot
(1\mbox{eV}/\langle m_{\nu}\rangle)
\ea
For this reason we neglected the SUSY induced $\Delta L = 0$ operators
in the effective Lagrangian \rf{effective}.

Now let us turn from the general consideration to the concrete case of
the SUSY contribution to $\znbb$-decay within the MSSM.
The following Lagrangian terms are relevant to $\znbb$-quark
transitions in Fig. 3(a)
\be{ltot}
{\cal L}_{int} = {\cal L}_{W{\bar f}f} + {\cal L}_{W{\tilde f}{\tilde f}}
           + {\cal L}_{\chi^+f{\tilde f}} +  {\cal L}_{\chi f{\tilde f}}
           + {\cal L}_{{\tilde g}{\tilde q}q}
           + {\cal L}_{W\chi^+\chi}.
\ee
The Lagrangian terms in the r.h.s. are explicitly given in
Appendix A.

Starting from this Lagrangian one can find 14 dominant diagrams
proportional to ${\tilde m}_M^2$ which contribute to the $\znbb$-quark
transition in Fig.3(a). They are listed in Fig.4. (Note, that in addition
to the graphs shown, there exist several graphs corresponding simply to
an exchange of two of the external momenta and are not shown for
brevity.) 
As seen, all diagrams in Fig.4 fall into 5 classes represented by
the last 5 diagrams  of the decomposition in Fig.3. The supersymmetric
part of these diagrams, as discussed above, can be parameterized by
the effective Lagrangian ${\cal L}_{\znbb}$ given in the general form of
Eq. \rf{effective}. One can reconstruct a specific form of this Lagrangian
in the MSSM comparing diagrams in Fig.3  and Fig.4 and separating the
basic SUSY vertices denoted in Fig.3 by the black spots representing
five different terms of the effective Lagrangian ${\cal L}_{\znbb}$ in
Eq. \rf{effective}. The next step of the derivation is based on the
standard approximate procedure relying on the fact that all intermediate
particles involved in evaluation of the effective SUSY vertices are heavy
SUSY particles with typical masses of order $m_{SUSY}$. As a result these
vertices can approximately be represented in the form of local operators.
The local form of the SUSY operators allows one to apply the Fierz
rearrangement and to collect the quark fields in the color-singlet
quark charged currents. Straightforward realization of this strategy
leads to an effective Lagrangian with the following leading order
operators violating the lepton number $L$ by 2 units
\ba{eff_MSSM}
{\cal L}_{\Delta L = 2} &=&
-  (\eta^{(1)}_{WW} + \eta^{(2)}_{WW} + \eta^{(3)}_{WW})\
\frac{W^-_{\mu} W^{-\ \mu}}{m_{SUSY}}\cdot
                              \bar e (1+\gamma_5) e^c  +  \\ \nn
&+&    (\eta_{\tilde g \tilde u} + \eta_{\tilde g \tilde d})\
\frac{j^{\mu}_{{}_{AV}}j_{\mu {}_{AV}}}{m_{SUSY}^5 }
\cdot \bar e  (1+\gamma_5) e^c.
\ea
The color-singlet quark charged currents are defined as usual
\ba{qcur}
j^{\mu}_{\mbox{\tiny AV}} =
       \cos\theta_c \ \bar u \gamma^{\mu}(1 - \gamma_5) d .
\ea
Note that since we take only the leading order contributions
into account in Eq. \rf{eff_MSSM} not all possible terms of the
decomposition Eq. \rf{effective} are retained in Eq. \rf{eff_MSSM}.
Naively one might have expected that the diagrams in Fig. 4 (and the
corresponding terms in Eq. \rf{effective}) are ordered with respect
to decreasing importance, since a larger number of heavy sparticles
in the loops results in larger (loop) suppression factors. However,
the explicit calculation shows that this is not the case. Terms
corresponding in structure to the 2nd, 3rd and the 5th terms in Eq.
\rf{effective} are suppressed by the helicity structure of the basic
MSSM interactions and/or typical suppression factors of order
$m_e/m_{SUSY}$ from left-right sfermion mixing or higgsino-like
interactions.

Retaining only leading contributions to the operator structures in
Eq. \rf{eff_MSSM}, the dimensionless lepton number violating parameters
take the form

\ba{etai}
\eta_{\tilde g \tilde d} &=&
\frac{g_s^2 g^4}{72}\Big(\frac{{\tilde m}_M}{m_{SUSY}}\Big)^2
                      \sum_{i,j} U_{i1}V_{i1}U_{j1}V_{j1}
                      \Big(\frac{m_{\chi^{\pm}_j}}{m_{SUSY}}\Big)
                      \Big(\frac{m_{\chi^{\pm}_i}}{m_{SUSY}}\Big)
                       \times \\ \nn
                 &\times&   \Big(\frac{m_{\tilde g}}{m_{SUSY}}\Big)
{\cal G}(m_{\tilde g}, m_{\chi^{\pm}_j}, m_{\chi^{\pm}_i}),  \\
\eta_{\tilde g \tilde u} &=&
\frac{g_s^2 g^4}{72}\Big(\frac{{\tilde m}_M}{m_{SUSY}}\Big)^2
                     \sum_{i,j} V^2_{i1}V^2_{j1}
                      \Big(\frac{m_{\tilde g}}{m_{SUSY}}\Big)
            {\cal F}(m_{\tilde g}, m_{\chi^{\pm}_j}, m_{\chi^{\pm}_i}), \\
\eta^{(1)}_{WW} &=&
\frac{g^4}{4}  \Big(\frac{{\tilde m}_M}{m_{SUSY}}\Big)^2
\sum_{i,j,k} V_{k1}V_{j1} \times \\ \nn
& \times & \Big[{\cal O}^L_{ik}{\cal O}^L_{ij}
\Big(\frac{m_{\chi_i}}{m_{SUSY}}\Big)
{\cal J}(m_{\chi_i},m_{\chi^{\pm}_j},m_{\chi^{\pm}_k}) \\ \nn
&+& {\cal O}^R_{ik}{\cal O}^L_{ij}
\Big(\frac{m_{\chi^{\pm}_k}}{m_{SUSY}}\Big)
{\cal J}(m_{\chi_i},m_{\chi^{\pm}_j},m_{\chi^{\pm}_k}) \\ \nn
& + & {\cal O}^R_{ik}{\cal O}^R_{ij}
\Big(\frac{m_{\chi^{\pm}_j}}{m_{SUSY}}\Big)
\Big(\frac{m_{\chi^{\pm}_k}}{m_{SUSY}}\Big)
\Big(\frac{m_{\chi_i}}{m_{SUSY}}\Big)
{\cal I}(m_{\chi_i},m_{\chi^{\pm}_j},m_{\chi^{\pm}_k})\Big], \\
\eta^{(2)}_{WW} &=&
\frac{g^4}{4}  \Big(\frac{{\tilde m}_M}{m_{SUSY}}\Big)^2
\sum_{i,j} {\cal J}(m_{\tilde e}, m_{\chi_i},
m_{\chi^{\pm}_j})\epsilon_{L_i}(e)V_{j1} \times \\ \nn
&\times& \left[{\cal O}^R_{ij}
                    \Big(\frac{m_{\chi^{\pm}_j}}{m_{SUSY}}\Big) +
{\cal O}^L_{ij} \Big(\frac{m_{\chi_i}}{m_{SUSY}}\Big)\right], \\
\label{(etaW)}
\eta^{(3)}_{WW} &=&
\frac{g^4}{4}  \Big(\frac{{\tilde m}_M}{m_{SUSY}}\Big)^2
\sum_{i} {\cal J}(m_{\chi_i}, m_{\tilde e}, m_{\tilde e})
\epsilon^2_{L_i}(e) \Big(\frac{m_{\chi_i}}{m_{SUSY}}\Big)
\ea
The dimensionless loop factors ${\cal F}(m_i), {\cal G}(m_i),
{\cal J}(m_i)$ and ${\cal I}(m_i)$ are given in Appendix B. They
depend on the sparticle masses $m_i$  in the corresponding loop.
Recall that $g$ and $g_s$ are the $SU(2)_L$ and $SU(3)_c$ coupling
constants. Further definitions on couplings and mixing parameters
can be found in Appendix A.

The next step of the calculation deals with reformulating the problem
in terms of nucleon degrees of freedom instead of quark ones. This is
relevant for the nuclear structure part of the calculations.

\section{From quark to nuclear level}

So far the discussion has focussed on particle physics aspects,
deriving the low-energy effective Lagrangian in Eq. \rf{eff_MSSM}
formulated in terms of quark fields. However our goal is the calculation
of the amplitude $R_{0\nu\beta\beta}$ of $0\nu\beta\beta$-decay which
is a nuclear process proceeding not at the level of quark degrees of
freedom but at the level of nucleon ones. Formally one can write down
\ba{DefL}
{\cal R}_{\znbb} \ &=& <(A, Z + 2), 2 e^-| S - 1|(A, Z)> = \\ \nn
&=& <(A, Z + 2), 2 e^-| T exp[i \int d^4 x {\cal L}_{\znbb}(x)] |(A, Z)>
\ea

where the effective Lagrangian
${\cal L}_{\znbb} =  {\cal L}_{W{\bar f}f} + {\cal L}_{\Delta L = 2} $
is given by Eqs. \rf{lcc}, \rf{eff_MSSM}.
The nuclear structure is involved via the initial (A,Z) and
the final (A, Z+2) nuclear states having the same atomic
weight A, but different electric charges Z and Z+2. 
The standard framework for the calculation of this nuclear matrix
element is the non-relativistic impulse approximation(NRIA). It
implies the substitution of the quark current $j^{\mu}{{}_{AV}}$ in
the effective Lagrangian ${\cal L}_{\znbb}$  in Eq. \rf{DefL} by
the non-relativistic nucleon current
$$
j^{\mu}{{}_{\mbox{\tiny AV}}} \stackrel{\mbox{\scriptsize NRIA}}
{\longrightarrow} J^{\mu} .
$$
The latter is an incoherent sum over individual nucleon currents of
a nucleus and is given by the formula \cite{NRL}
\ba{non-rel}
J^{\mu} ({\bf x})  &=&
\sum_i \tau_{+}^{(i)} \left[(f_V - f_A C_i) g^{\mu 0} -
(f_A \sigma^k_i + f_V D^k_i) g^{\mu k} \right] \frac{m_A^3}{8\pi}
e^{-m_A|{\bf x} - \bf{r}_i|},
\ea
Here $f_V\approx 1,\ f_A\approx 1.261$, $g^{\mu\nu}$ is the metric
tensor, $\tau_{+}^{(i)}$ is the isospin raising operator, ${\bf r}_i$
is the position of the $i$th  nucleon, the superscript $k$ stands for
the spatial component. $C_i$ and ${\bf D}_i$ are the well-known scalar
and the vector nuclear recoil terms given by \cite{doi85}
\ba{recoil}
C_i &=& \frac{1}{2 m_P}
\left[({\bf p}_i + {\bf p}'_i)\cdot \si\  - \
\frac{f_P}{f_A} (E_i - E'_i) \ {\bf q}_i\cdot \si\right] \\
{\bf D}_i &=& \frac{1}{2 m_P}
\left[({\bf p}_i + {\bf p}'_i)\ \  + \ \
i (1 - 2 m_P\frac{f_W}{f_V}) {\bf q}_i\times\si\right],
\ea
Here $({\bf p}_i, E_i)$ and $({\bf p}'_i, E'_i)$  are initial and
final 3-momentum and energy of  the $i$th nucleon and the 3-momentum
transfer is ${\bf q}_i = {\bf p}_i - {\bf p}'_i$. The nucleon couplings
obey the relations
\ba{nucl_coupl}
f_W/f_V =  - (\mu_p - \mu_n)/(2 m_P) \approx 3.7/(2 m_P), \ \ \
f_P/f_A = 2 m_P/m_{\pi}^2,
\ea
where $m_{\pi}$ is the pion mass and $\mu_{p(n)}$ is the proton (neutron)
magnetic moment.

Since we are interested in the dominant contributions only, recoil
terms will be neglected in the rest of this paper and have been given
above for completeness.

The exponential factor in Eq. \rf{non-rel}  is introduced instead of
the local delta function in order to take into account the finite nucleon
size. It  is the Fourier transform of the nucleon form factor
$F({\bf q}^{\ 2})$ in the
conventional parameterization a dipole form
\ba{dip}
F({\bf q}^2) = \left(1 + \frac{{\bf q}^2}{m_A^2} \right)^{-2}
\ea
with $m_A = 0.85$GeV. The finite nucleon size effects are known
\cite{Vergados2} to be important for the short-distance contributions
to the $\znbb$-amplitude such as those corresponding to the dominant
terms in the effective Lagrangian \rf{eff_MSSM}.

Now, starting from Eq. \rf{DefL}, it is straightforward to calculate
the $\znbb$-amplitude $R_{\znbb}$  within the non-relativistic
impulse approximation. The final result for the $0^+\rightarrow 0^+$
transition amplitude can be written as follows
\ba{sep}
R_{\znbb}(0^+\rightarrow 0^+) &=&
 \frac{\eta^{SUSY}}{m^5_{SUSY}}
\sqrt{2}\cdot C^{-1}_{0\nu}\Big[{\bar e}(1+\gamma_5)e^c\Big]
                <F|\Omega^{SUSY}|I>.
\ea
The normalization factor is
\ba{norm}
C_{0\nu} = \frac{4 \pi}{m_P m_e} \frac{R_0}{f_A^2}.
\ea
Here, $R_0$ is the nuclear radius, $m_P$ and $m_e$ are the proton
and the electron masses.

The effective lepton number violating parameter is defined as
\ba{param_eff}
\eta^{SUSY} &=&  (\eta_{\tilde g \tilde d} +
\eta_{\tilde g \tilde u}) +
g^2 \Big(\frac{m_{SUSY}}{M_W}\Big)^4
(\eta^{(1)}_{WW} + \eta^{(2)}_{WW}+ \eta^{(3)}_{WW}).
\ea

In Eq. \rf{sep} we have introduced the transition operator $\Omega^{SUSY}$
in order to separate the nuclear physics part of the calculation from
the particle physics one. Having the transition operator one can
calculate the corresponding nuclear matrix element for any
$\znbb$-decaying candidate isotope within any specific model of
nuclear structure. From now on  we define
\ba{defm1}
{\cal M}^{SUSY} = <F|\Omega^{SUSY}|I> \\ \nn
\ea

This nuclear matrix element is found to be equal to those
for heavy neutrino exchange \cite{lrmodel}
\be{defm3}
{\cal M}^{SUSY} = \Big\{ {\cal M}_{F,N}
                   - {\cal M}_{GT,N} \Big\},
\ee
where

\be{mgth}
{\cal M}_{GT,N} = \Big(\frac{m_A^2}{m_e m_p}\Big)
                 \big< F |\Omega_{GT,N} | I \big>
\ee
where
\be{omgth}
\big< F |\Omega_{GT,N} | I \big> =
\big< F | \sum_{i \ne j} \tau_{+}^{(i)} \tau_{+}^{(j)}
                     \si \cdot \sj
                     \left(\frac{R_0}{r_{ij}}\right)
                      F_{N}(x_{A})                | I \big>
\ee
and
\be{mfh}
{\cal M}_{F,N} = \Big(\frac{m_A^2}{m_e m_p}\Big)(\frac{f_V}{f_A})^2
                 \big< F |\Omega_{F,N} | I \big>
\ee
where
\be{omfh}
\big< F |\Omega_{F,N} | I \big> =
                 \big< F | \sum_{i \ne j} \tau_{+}^{(i)} \tau_{+}^{(j)}
                     \left(\frac{R_0}{r_{ij}}\right)
                      F_{N}(x_{A})                | I \big>.
\ee
Here, $F_N(x_A)$ is the short-ranged potential
\ba{F_N}
F_N (x_A) &=& 4 \pi m_{A}^{6}  r_{ij} \int
\frac{d^3{\bf q}}{(2\pi)^3} \frac{1}{(m_A^2 + {\bf q}^2)^4}
        e^{i {\bf q} {\bf r}_{ij}}
\ea
with $x_A = m_A r_{ij}$.
This potential takes into account the finite nucleon size, see
Eq. \rf{dip}, its analytic solution is given by the formula
\ba{Fnn}
F_N(x)  = \frac{x}{48} (3 + 3 x + x^2) e^{-x}. 
\ea
The above definitions are general in the sense that one can apply
nuclear wave functions of any nuclear structure model for their
calculation. For the following analysis, on the other hand, numerical
values for the matrix elements are needed.

In our numerical analysis we will use the following value for the
$\znbb$ decay of $^{76}$Ge \cite{lrmodel}
\ba{num}
{\cal M}^{SUSY} = 289 .
\ea
This value is based on a pn-QRPA model \cite{mut89}, which has been
discussed already several times in the literature \cite{mut89} and has
been applied previously to calculations of the R-parity violating
contributions to $\znbb$ decay \cite{HKK,HKK1}, as well as to $\znbb$
decay in left-right symmetric models \cite{lrmodel}. We will
therefore not repeat the details of the calculation here, and refer
for brevity to \cite{mut89}. Uncertainties associated with the pn-QRPA
have been discussed in \cite{HKK} for ${\cal M}_{F,N}$ and
${\cal M}_{GT,N}$.

\section{$\znbb$ constraints on (B-L)-violating Sneutrino Mass}

The $\znbb$-decay amplitude given in Eq. \rf{sep} leads to the
following half-life formula
\ba{hl-theory}
\big[ T_{1/2}^{{0\nu\beta\beta}}(0^+ \rightarrow 0^+) \big]^{-1} =
 G_{01} \frac{4 m_P^2}{G_F^4}
\left|\frac{\eta^{SUSY}}{m_{SUSY}^5} {\cal M}^{SUSY}\right|^2.
\label{half-life}
\ea
Here $G_{01}$ is the standard phase space factor tabulated
for various nuclei in  \cite{doi85}.

Eq. \rf{hl-theory} takes into account only the contributions from
sneutrino exchange. There might be other contributions which we assume
not to cancel the SUSY contribution. If there is no unnatural fine tuning
between different contributions we may retain only the SUSY one in deriving
upper bounds for the lepton number violating parameters.

The most stringent experimental lower limit on $0\nu\beta\beta$-decay
has been obtained for $^{76}$Ge \cite{hdmo97}
\ba{HD}
T_{1/2}^{{0\nu\beta\beta}-exp}(0^+ \rightarrow 0^+)
\hskip2mm \geq \hskip2mm
1.0 \times 10^{25}  years\ \ \  90 \% \ \mbox{c.l.}
\ea
Combining this bound with Eq. \rf{hl-theory} and the numerical value of
the nuclear matrix element ${\cal M}^{SUSY}$ given in Eq. \rf{num} we get
the following constraint on the effective MSSM parameter
\ba{sum}
\eta^{SUSY} \leq 1.0 \times 10^{-8}
            \Big(\frac{m_{SUSY}}{\rm 100 GeV}\Big)^5
\ea
Since we are interested in deriving constraints on the (B-L)-violating
sneutrino mass $\tilde m_M$ from $\eta^{SUSY}$, we will adopt the following
simplifying assumptions. Assume all SUSY particle masses to be equal to the
effective SUSY breaking scale $m_{SUSY}$ introduced in Eq. \rf{effective}
and consider two limiting cases for the lightest neutralino $\chi$
composition. In the first case it is assumed to be a pure B-ino as
suggested by the SUSY solution of the dark matter problem \cite{Jungman},
in the second a pure Higgsino. These two cases can be understood as
extreme cases, and actual values for ${\tilde m}_M$ for other choices
of the neutralino composition should therefore lie in between the two
extreme values given below. In the Higgsino case essentially only the last
three graphs in Fig.4 with gluino lines contribute to $\znbb$-decay.
As seen from Eq. \rf{etai} the corresponding lepton number violating
parameter $\eta_{\tilde q}$ does not depend on the neutralino composition
and survives in this limiting case.
With the currently accepted \cite{PDG} values of the gauge coupling
constants and W-boson mass we derive
\ba{lim_znbb}
{\tilde m}_M &\leq& 2
\Big(\frac{m_{SUSY}}{100\mbox{GeV}}\Big)^{3/2}\mbox{GeV} ,
\ \ \ \chi \ \sim \
\tilde{B}, \\
{\tilde m}_M &\leq&
11
\Big(\frac{m_{SUSY}}{100\mbox{GeV}}\Big)^{7/2}\mbox{GeV} ,
\ \ \ \chi \ \sim \ \tilde{H}.
\ea

\section{Neutrino mass constraints}

As already mentioned, the sneutrino contributes to the Majorana neutrino
mass $m_M^{\nu}$ at the 1-loop level via the (B-L)-violating propagator
Eq. \rf{propagator_M} proportional to the sneutrino (B-L)-violating
mass parameter $\tilde m_M^2$. The corresponding diagram given
in Fig. 2(f) gives rise to an induced Majorana neutrino mass
$\delta m^{\nu}$. Thus, in the presence of a non-zero
tree-level contribution $m_M^{\nu(tree)}$ the total neutrino mass is
\ba{tot}
m_M^{\nu} =  m_M^{\nu(tree)} + \delta m^{\nu}. 
\ea
As seen from Eq. \rf{propagator_M} the (B-L)-violating sneutrino
propagator in momentum space is a rapidly decreasing function of
momentum q. In the ultraviolet limit it behaves as
$
\Delta_{\tilde\nu}^M(q) \sim 1/q^4
$
unlike an ordinary scalar propagator decreasing only as $\sim 1/q^2$.
As a result, the 1-loop diagram in Fig. 2(f) leads to a finite loop
integral. Hence, the sneutrino induced  Majorana neutrino mass
$\delta m^{\nu}$ is a calculable object. It is given by the formula
\ba{matrix}
\delta m^{\nu_{(i)}} =  g^2 \sum_{k=1}^{4} \epsilon^{\nu^2}_{L_k}  m_{\chi_k}
\tilde m_{M_{(i)}}^2 I(\tilde m_M^2, \tilde m_D^2, m_{\chi_k})
\ea
where the subscript $i$ stands for generation. 
The neutralino-neutrino-sneutrino coupling is
$\epsilon^{\nu}_{L_k} = \tan\theta_W {\cal N}_{k1} - {\cal N}_{k2}$.
The loop integral
is defined as
\ba{integral}
I(\tilde m_M^2, \tilde m_D^2, m_{\chi_k}) =
- i \int\frac{d^4q}{(2\pi)^4}
\frac{1}{(\tilde m_1^2 - q^2)}\frac{1}{(\tilde m_2^2 - q^2)}
\frac{1}{(m_{\chi_k}^2 - q^2)}. 
\ea
For an approximate numerical estimation we take all superpartner masses
equal to the common mass scale of supersymmetry breaking
$
m_{\chi_k} \approx \tilde m_1 \approx \tilde m_2 \approx
m_{SUSY}.
$
Then in this approximation  one gets for the lightest neutralino $\chi$
contribution the following constraint on the sneutrino mass splitting
parameter
\ba{1-loop-nu}
\tilde m_{M(i)}  \leq \frac{1.7\cdot 10^{-2}}{|\tan\theta_W {\cal N}_{11} -
{\cal N}_{12}|}
\left(\frac{m_{SUSY}}{100\mbox{GeV}}\right)^{1/2}
\left(\frac{m_{\nu_{(i)}}^{exp}}{1\mbox{eV}}\right)^{1/2} {\rm GeV}.
\ea
Here $m_{\nu_{(i)}}\leq m_{\nu_{(i)}}^{exp}$ are the best laboratory
limits on the neutrino masses which can be summarized as \cite{PDG}
$m_{\nu_{(e)}}^{exp} = 15\mbox{eV}, \ m_{\nu_{(\mu)}}^{exp} =
170\mbox{KeV},\  m_{\nu_{(\tau)}}^{exp} =  23\mbox{MeV}$. 
Eq. \rf{1-loop-nu} assumes that there is no significant cancellation
in Eq. \rf{tot} between the tree and 1-loop level contributions.

If the neutralino is B-ino dominant, than we derive the following limits on
the sneutrino mass splitting parameter
\ba{li}
\tilde m_{M(e)} \leq 120\mbox{MeV},\ \ \
\tilde m_{M(\mu)} \leq 13\mbox{GeV}, \ \ \
\tilde m_{M(\tau)} \leq 149\mbox{GeV}.
\ea
Thus, for the second and third generation sneutrinos large splittings
are not excluded by experimental data.

An interesting question to ask is whether there are certain loopholes
in the constraints on the sneutrino mass splitting derived from the
experimental upper limits on neutrino masses. It could happen, for
instance that the lightest neutralino is higgsino dominated, in which
case one would expect that the bound \rf{1-loop-nu} might have to
be relaxed considerably. However, one should remember that all
neutralino states contribute to \rf{matrix}, so that even if there
is no constraint from the lightest neutralino the other mass
eigenstates will provide a finite contribution to the neutrino mass.

In order to investigate this question a little bit more quantitatively
we did a numerical scan of the SUSY parameter space, calculating
upper bounds on ${\tilde m}_M$ taking into account all four
neutralino states.  In this case instead of Eq. \rf{1-loop-nu} we have
\ba{1-loop-nu-compl}
\tilde m_{M(i)}  \leq \frac{1.7\cdot 10^{-2}{\rm GeV}}
{(\sum_{i=1}^{4}(\tan\theta_W N_{i1} - N_{i2})^2 y_i
          C_3(x,y_i))^{1/2}}
\left(\frac{m_{SUSY}}{100\mbox{GeV}}\right)
\left(\frac{m_{\nu_{(i)}}^{exp}}{1\mbox{eV}}\right)^{1/2} .
\ea
Here $x={\tilde m}_D/m_{SUSY}$ and
$y_i = m_{\chi_i}/m_{SUSY}$. $C_3(x,y_i)$ takes into
account the fact that the masses of the particles in the loop
are no longer taken to be equal. In the limit where
${\tilde m}_M \ll {\tilde m}_D$ the 1-loop integral $C_3(x,y_i)$ is given by
\ba{defc3}
C_3(x,y) = 2 \frac{x^2+y^2({\rm Log}(y^2)-{\rm Log}(x^2)-1)}
                  {(y^2-x^2)^2}
\ea
$C_3(x,y)$ is normalized such that it approaches $1$ in the limit
where $x$ and $y$ approach $1$.
We let the parameters of the neutralino mass
matrix vary from ($0-1000$) GeV for $\mu$ and $M_2$, $\tan\beta$
from ($1-50$), for both positive and negative $\mu$. The unification
condition $M_1 = (5/3)\tan^2\theta_W M_2$ has been assumed in this
calculation. We required the lightest mass eigenstate to be heavier
than $20$ GeV, motivated by the LEP measurements. About $10^8$
solutions were calculated. From these we calculated the ``average
constraint'' and an ``absolute'' upper bound. These are:
\ba{statistics}
\tilde m_{M(i)}  \leq 60 (125)
\left(\frac{m_{\nu_{(i)}}^{exp}}{1\mbox{eV}}\right)^{1/2} {\rm MeV}.
\ea
on average (``absolute''), if ${\tilde m_D}\approx m_{SUSY} = 100$ GeV is
assumed. These numbers are about a factor of $2$ ($4$) less stringent than
taking only the lightest neutralino (being bino) fixed at $100$ GeV.
This simply reflects the fact that within the above-mentioned parameter
ranges many solutions exist where even the lightest neutralino mass state
can be considerably heavier than $100$ GeV. On the other hand, it
seems that within the typical range of SUSY parameters, the
constraint on ${\tilde m}_M$ is essentially ``stable'' and has
to be taken seriously. (This conclusion remains unchanged even if
we drop the unification assumption on $M_1$, although the bounds
might have to be slightly relaxed in some cases.)

Note that, in principle, more stringent limits on $\tilde m_M$ than
in Eq. \rf{li} could be derived from the upper bounds on the neutrino mass
given by non-observation of  $\znbb$ decay.
However, in this case the situation is more complex,
since $\znbb$ decay measures an effective neutrino mass
$\langle m^{\nu}_M \rangle = \sum^{\prime}
U^2_{ej} m_j$, where $U_{ej}$ are mixing coefficients connecting the
weak and the mass eigenstate basis for neutrinos. Thus limits on
${\tilde m}_M$ derived from the neutrino mass limit of $\znbb$ decay
will also depend on assumptions on neutrino mixing. If one assumes
for simplicity $U_{ej} \approx \delta_{e1}$ one could derive
$\tilde m_{M(e)} \leq 22$ MeV from the data on $^{76}$Ge
\cite{hdmo97}.

\section{Conclusion}
In summary, we have proven a low-energy theorem for weak scale softly
broken supersymmetry relating the (B-L)-violating mass terms of the
neutrino and the sneutrino as well as the amplitude of neutrinoless
double beta decay. This theorem can be regarded as a supersymmetric
generalization of the well-known theorem \cite{SV} relating only the
neutrino Majorana mass and the neutrinoless double beta decay
amplitude.

According to Eq. \rf{snu_DM} the parameter $\tilde m_M$ describes a
splitting in the sneutrino mass spectrum. This splitting leads to
mixing in the sneutrino-antisneutrino ($\tilde\nu-\tilde\nu^c$)
system and to the effect of lepton number violating
$\tilde\nu-\tilde\nu^c$ oscillations \cite{sn_accel},
\cite{Grossman_Haber}.

The mass splitting parameter $\tilde m_M$ is constrained by
the experimental data on neutrinoless double beta decay $\znbb$
and the neutrino mass discussed in the present paper. The neutrino
mass constraint on $\tilde m_M$ is found to be more stringent then the
direct $\znbb$-decay constraint. This is opposite to the conclusion
reached for R-parity violating SUSY, for which the direct double beta
decay constraints have been found to be more stringent than those
derivable from the neutrino mass \cite{HKK_top,HKK}. 
However, in contrast to the neutrino mass limits, the corresponding 
constraint on $\tilde m_M$ from neutrinoless double beta decay 
is completely independent of assumptions about neutralino masses 
and mixings. 

The constraints derived here, nevertheless, leave quite some room for
accelerator searches for sneutrino mediated (B-L)-violating effects
for the second and third generation. The sneutrino mass splitting
parameter $\tilde m_M$ might be searched for at future colliders such
as the NLC  or a first muon collider \cite{sn_accel}, \cite{Grossman_Haber}.
Probably, dedicated searches for Majorana sneutrinos have a chance of
detecting positive signal, within the above discussed low-energy limits.
For ${\tilde \nu}_e$, on the other hand, these limits seem to be too
stringent and accelerator experiments should not be expected
to give positive signals.

\vskip10mm
\centerline{\bf Acknowledgments}

\bigskip
We thank V.A.~Bednyakov,  for helpful discussions.
The research described in this publication was made possible in part
(S.G.K.) by Grant GNTP 315 NUCLON from the Russian ministry of science.
M.H. would like to thank the Deutsche Forschungsgemeinschaft
for financial support by grants kl 253/8-2 and 446 JAP-113/101/0.

\section{Appendix A. Supersymmetric Lagrangian terms contributing to
                     the $\znbb$-decay amplitude}

In the presence of the (B-L)-violating (s)neutrino masses,
given in Eq. \rf{complete}, $\znbb$-decay is triggered by the following
terms of  the MSSM  Lagrangian

\be{ltotapp}
{\cal L}_{MSSM} = {\cal L}_{W{\bar f}f} + {\cal L}_{W{\tilde f}{\tilde f}}
           + {\cal L}_{\chi^+f{\tilde f}} +  {\cal L}_{\chi f{\tilde f}}
           + {\cal L}_{{\tilde g}{\tilde q}q}
           + {\cal L}_{W\chi^+\chi}
\ee
The individual terms can be found in the standard sources like
ref. \cite{Haber}, \cite{gun86}. Let us list them explicitly.
\medskip

{\bf a)} {\it Gauge boson-fermion-fermion}

\medskip
\noindent
This is nothing but the usual standard model charged-current
Lagrangian,

\ba{lcc}
{\cal L}_{W{\bar f}f} & = & - \frac{g}{\sqrt{2}}
\left[W_{\mu}^{+}(\bar u_L \gamma^{\mu} d_L + \bar{\nu}_L \gamma^{\mu} e_L) +
{\rm h.c.}\right] \\
&\equiv  & {\cal L}_{W{\bar q}q} + {\cal L}_{W{\bar l}l}.
\ea

\medskip
{\bf b)} {\it Gauge boson-sfermion-sfermion}

\medskip
\noindent
Only the charged-current part of this type of the MSSM interactions
is of interest in $\znbb$ decay:
\ba{lwsfsf}
{\cal L}_{W{\tilde f}{\tilde f}} & \equiv &
{\cal L}_{W{\tilde q}{\tilde q}} + {\cal L}_{W{\tilde l}{\tilde l}} \\
& = & - \frac{ig}{\sqrt{2}}W_{\mu}^{+}
\big( {\tilde u}^*_L {\olrap}^{\mu}{\tilde d}_L\big)
- \frac{ig}{\sqrt{2}}W_{\mu}^{+}
\big( {\tilde \nu}^*_L {\olrap}^{\mu}{\tilde e}_L\big)
+ {\rm h.c.}
\ea
Note that ${\olrap}^{\mu}$ is defined by
${\olrap}^{\mu} = {\overleftarrow \partial}^{\mu}
- {\overrightarrow \partial}^{\mu}$.

\medskip
{\bf c)} {\it Chargino-fermion-sfermion}

\medskip
\ba{lagw}
{\cal L}_{\chi^+ f{\tilde f}} &=&
       C^u _{LL} \cdot {\bar u_L}\chi^+_i {\tilde d}_L
     + C^d_{LL} \cdot {\bar d_L}\chi^{c+}_i {\tilde u}_L +
     C^u_{LR} \cdot {\bar u_L}\chi^+_i {\tilde d}_R + \\ \nn
    &+& C^u_{RL} \cdot {\bar u_R}\chi^+_i {\tilde d}_L
    + C^d_{RL} \cdot {\bar d_R}\chi^-_i {\tilde u}_L +
     C^d_{LR} \cdot {\bar d_L}\chi^-_i {\tilde u}_R +\\ \nn
 &+&   C^{\nu}_{LL} \cdot {\bar \nu_L}\chi^+_i {\tilde e}_L
     + C^e_{LL} \cdot {\bar e_L}\chi^-_i {\tilde \nu}_L  +
{\rm h.c.},
\ea
where the following shorthands have been defined:
\ba{cwh}\nn
C^{u, \nu}_{LL} & = & -g U_{i1}, \ \ \
C^{d, e}_{LL} = -g V_{i1} , \\ \nn
C^u_{LR} & = & \frac{g m_d}{\sqrt{2}m_W cos\beta} U_{i2}, \ \ \
C^u_{RL} = \frac{g m_u}{\sqrt{2}m_W sin\beta} V^*_{i2}, \\ \nn
C^d_{RL} & = & \frac{g m_d}{\sqrt{2}m_W cos\beta} U^*_{i2}, \ \ \
C^d_{LR} = \frac{g m_u}{\sqrt{2}m_W sin\beta} V_{i2}.
\ea
Coefficients $C_{LL}, C_{RR}$ are fermion-sfermion couplings to the
gaugino component of the chargino while $C_{LR}, C_{RL}$ describe
couplings to the Higgsino component. The latter are proportional to
the fermion mass and, therefore, can be neglected in the
fermion-sfermion sector as is done in the present paper.

The chargino mixing matrices $U_{ij}$ and $V_{ij}$ are defined from
the diagonalization of the chargino mass matrix $M_{\chi^{\pm}}$
\ba{charg_mix}
U^*\cdot M_{\chi^{\pm}} V^{\dagger} = Diag\{m_{\chi^{\pm}} \},
\ea
For details see Ref. \cite{gun86}.

\medskip
{\bf d)} {\it Neutralino-fermion-sfermion}

\medskip
\noindent
The neutralino interaction can be written as
\ba{lagn}
{\cal L}_{\chi f{\tilde f}} &=& - \sqrt{2} g \
    \big[\epsilon_{L(i)}^{\psi} {\bar \psi_L}\chi_i {\tilde \psi}_L
     + \epsilon_{R(i)}^{\psi} {\bar \psi_R}\chi_i {\tilde \psi}_R
+\\ \nn
&+& \epsilon_{LR(i)}^{\psi} {\bar \psi_L}\chi_i {\tilde \psi}_R +
\epsilon_{RL(i)}^{\psi} {\bar \psi_R}\chi_i {\tilde \psi}_L
\big]  - {\rm h.c.},
\ea
where $\psi_L = u_L, d_L, e_L, \nu_L$ and their scalar superpartners
$\tilde\psi_L = \tilde u_L, \tilde d_L, \tilde e_L,\tilde\nu_L$.
The corresponding chiral coefficients are
\ba{defeps} \nn
\epsilon_{L(i)}^{\psi} &=& - T_3(\psi) {\cal N}_{i2} +
                       {\rm tan}\theta_W[T_3(\psi)-Q(\psi)]{\cal N}_{i1}, \\
\epsilon_{R(i)}^{\psi} &=& Q(\psi){\rm tan}\theta_W{\cal N}_{i1},\\
\epsilon_{LR(i)}^{u} &=& \frac{m_{u}}{m_W sin\beta}{\cal N}_{j4}^*,\\
\epsilon_{RL(i)}^{u} &=& \frac{m_{u}}{m_W sin\beta}{\cal N}_{j4}, \\
\epsilon_{LR(i)}^{d} &=& \frac{m_{d}}{m_W cos\beta}{\cal N}_{j3}^*,\\
\epsilon_{RL(i)}^{d} &=& \frac{m_{d}}{m_W cos\beta}{\cal N}_{j3}.
\ea

Coefficients $N_{ij}$ are elements of the orthogonal neutralino mixing
matrix diagonalizing the neutralino mass matrix $M_{\chi}$. In the \rp
MSSM the neutralino mass matrix is identical to the MSSM one \cite{Haber}
and in the basis of fields
($\tilde{B}, \tilde{W}^{3}, \tilde{H}_{1}^{0},  \tilde{H}_{2}^{0}$)
has the form:
\be{MassM}  
                M_{\chi} =  \left(
                        \begin{array}{cccc}
 M_1 & 0 & -M_Z s^{}_W c_\beta & M_Z s^{}_W s_\beta \\
 0   & M_2 & M_Z c^{}_W c_\beta & -M_Z c^{}_W s_\beta\\
 -M_Z s^{}_W c_\beta &  M_Z c^{}_W c_\beta & 0 & -\mu  \\
  M_Z s^{}_W s_\beta & -M_Z c^{}_W s_\beta & -\mu & 0 \\
 \end{array}
                     \right),
\ee
where $c^{}_W = \cos\theta_W$, $s^{}_W = \sin\theta_W$,
$t^{}_W = \tan \theta_W$, $s_\beta = \sin\beta$, $c_\beta = \cos\beta$.
The angle $\beta$ is defined as $\tan\beta = <H_{2}^{0}>/<H_{1}^{0}>$.
Here $ <H_{2}^{0}>$ and $<H_{1}^{0}>$ are vacuum expectation values
of the neutral components $H_{2}^{0}$ and $H_{1}^{0}$ of the Higgs
doublet fields with weak hypercharges $Y(H_{2}^0)  = +1$ and
$Y(H_{1}^{0}) = -1$, respectively. The mass parameters $M_1, M_2$ are
the soft $SU(2)_L$ and $U(1)_Y$ gaugino masses. In grand unification
scenarios they are related to each other as follows
\ba{GUT}
M_1 = \frac{5}{3} \tan^2\theta_W \cdot M_2
\ea
By diagonalizing the mass matrix \rf{MassM} one can obtain four
neutralinos $\chi_i$ with masses $m_{\chi_i}$ and the field content
\be{admix}
\chi_i = {\cal N}_{i1} \tilde{B} +  {\cal N}_{i2}  \tilde{W}^{3} +
{\cal N}_{i3} \tilde{H}_{1}^{0} + {\cal N}_{i4} \tilde{H}_{2}^{0}.
\ee
Recall again that we use notations $\tilde{W}^{3}$, $\tilde{B}$ for
neutral $SU(2)_L \times U(1)$ gauginos and  $\tilde{H}_{1}^{0}$,
$\tilde{H}_{2}^{0}$ for higgsinos which are the superpartners of the
two neutral Higgs boson fields $H_1^0$ and $H_2^0$.

We apply a diagonalization by means of a real orthogonal matrix
${\cal N}$. Therefore the coefficients ${\cal N}_{ij}$ are real and
masses $m_{\chi_i}$ are either positive or negative. The sign of the
mass coincides with the CP-parity of the corresponding neutralino mass
eigenstate $\chi_i$. If necessary, a negative mass can be always made
positive by a redefinition  \cite{gun86} of  the neutralino field $\chi_i$.
It leads to a redefinition of the relevant mixing coefficients
${\cal N}_{ij} \rightarrow i\cdot {\cal N}_{ij}$.

\medskip
{\bf e)} {\it Gluino-squark-quark}

\medskip
\noindent
The gluino interaction is given by
\ba{gluino}
                {\cal L}_{\tilde g} = - \sqrt{2} g_3
        \frac{{\bf \lambda}^{(a)}_{\alpha \beta}}{2}
        \left( \bar q_L^{\alpha} \tilde g^{(a)} \tilde q_L^{\beta}
                 - \bar q_R^{\alpha} \tilde g^{(a)} \tilde q_R^{\beta}
                        \right)
			+ h.c.,
\ea
Here ${\bf \lambda}^{(a)}$ are $3\times 3$ Gell-Mann matrices
($a = 1,..., 8$). Superscripts $\alpha, \beta$ in eq. \rf{gluino} are
color indices.

\medskip
{\bf f)} {\it Gauge boson-chargino-neutralino}

\medskip
\noindent
In the notation of ref. \cite{Haber}

\be{lgbcn}
{\cal L}_{W\chi^+\chi} = g W_{\mu}^-{\bar \chi_i} \gamma^{\mu}
\Big( {\cal O}^L_{ij}P_L + {\cal O}^R_{ij}P_R \Big)
\chi^+_j  + {\rm h.c.},
\ee
where

\ba{doij}
{\cal O}^L_{ij} & = & - \frac{1}{\sqrt{2}} {\cal N}_{i4}V_{j2}^*
                      + {\cal N}_{i2}V_{j1}^* \\
{\cal O}^R_{ij} & = & + \frac{1}{\sqrt{2}} {\cal N}_{i3}^* U_{j2}
                      + {\cal N}_{i2}^* U_{j1}.
\ea
Further details and conventions on the definitions used can be found
in the paper by Haber and Kane \cite{Haber}.

\section{Appendix B. Box Integrals}

In this appendix some relevant formulas for the calculation of the
loop integrals are summarized. There are four kinds of integrals
encountered in the Eqs. \rf{etai}-\rf{etaW}:
\ba{intg}
&{\cal G}(m_1, m_2, m_3)&  =
  - i    \int \frac{d^4k}{(2\pi)^4}  \\ \nn
& \times &\frac{m_{SUSY}^{10}}{({\tilde m}_1^2- k^2)({\tilde m}_2^2 - k^2)
    (m_{\tilde d}^2 - k^2)^2(m_1^2 - k^2) (m_2^2 - k^2) (m_3^2 - k^2)},
\ea
\ba{intf}
&{\cal F}(m_1, m_2, m_3)&  =
  - i   \int \frac{d^4k}{(2\pi)^4} \\ \nn
&\times & \frac{m_{SUSY}^8 k^2}
          {({\tilde m}_1^2- k^2)({\tilde m}_2^2 - k^2)
    (m_{\tilde u}^2 - k^2)^2(m_1^2 - k^2) (m_2^2 - k^2) (m_3^2 - k^2)}
\ea
\ba{inti}
&{\cal I}(m_1, m_2, m_3) & =
  - i    \int \frac{d^4k}{(2\pi)^4} \\ \nn
&\times & \frac{m_{SUSY}^6}
          {({\tilde m}_1^2 - k^2)({\tilde m}_2^2 - k^2)
           (m_1^2 - k^2)(m_2^2 - k^2)(m_3^2 - k^2)}
\ea

\ba{intj}
&{\cal J}(m_1, m_2, m_3) & =
  - i    \int \frac{d^4k}{(2\pi)^4} \\ \nn
&\times & \frac{m_{SUSY}^4 k^2}
          {({\tilde m}_1^2 - k^2)({\tilde m}_2^2 - k^2)
           (m_1^2 - k^2)(m_2^2 - k^2)(m_3^2 - k^2)}
\ea

All four integrals are finite and can be calculated analytically using
standard methods. Simple solutions can be found for the case when the
masses of all particles in the loops are assumed to be equal to
$m_{SUSY}$. In this limit one finds
\ba{GFIJ}
{\cal G}(m_{SUSY}) &=& (480\pi^2)^{-1}, \ \
{\cal F}(m_{SUSY}) = (960\pi^2)^{-1}, \\ \nn
{\cal I}(m_{SUSY}) &=& {\cal J}(m_{SUSY}) = (192\pi^2)^{-1}.
\ea

\newpage



\bigskip
{\large\bf Figure Captions}

\bigskip
\noindent
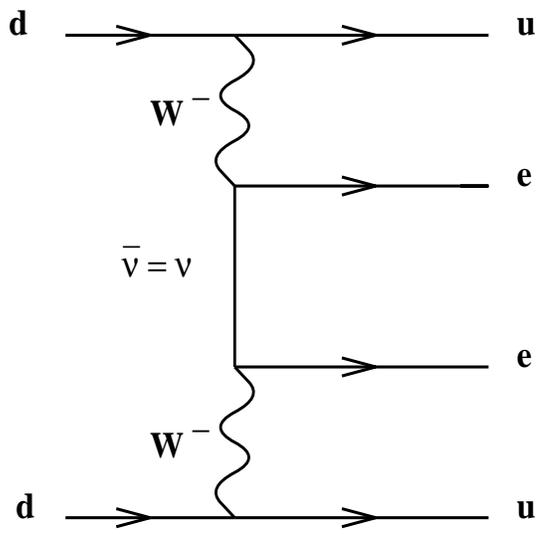
{\bf Fig.1.:} (a) the conventional Majorana neutrino mass mechanism of
$\znbb$-decay; (b) an  example of a $R_P$ conserving SUSY contribution
to the $\znbb$-decay.

\bigskip
\noindent
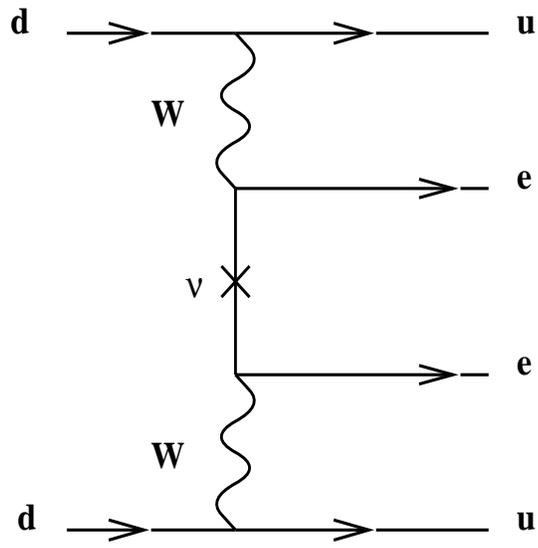
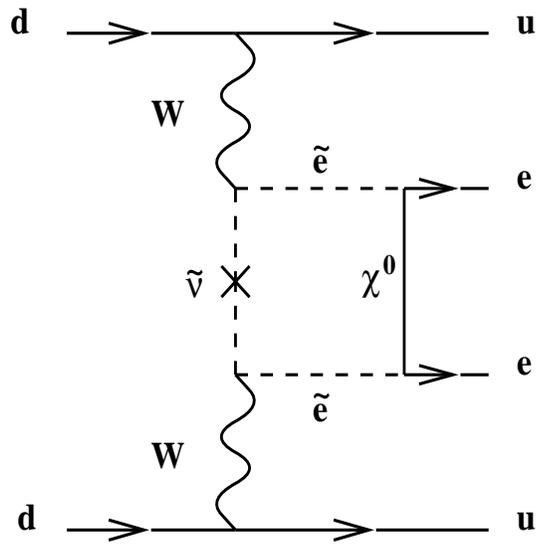
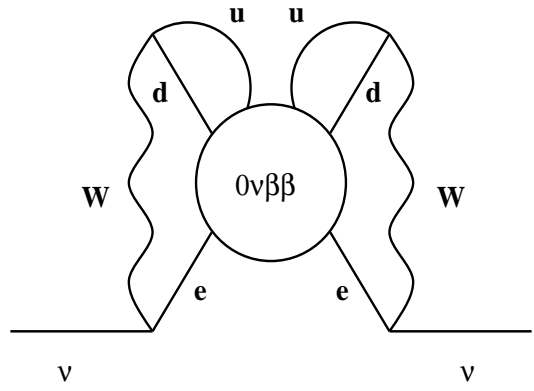
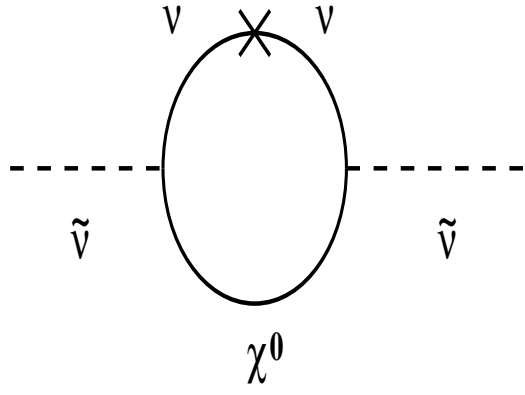
{\bf Fig.2.:} Lowest order perturbation theory diagrams representing
the relation between the neutrino Majorana mass $m_M^{\nu}$, the
"Majorana"-like (B-L)-violating sneutrino mass $\tilde m_M$ and the
amplitude of neutrinoless double beta decay $R_{0\nu\beta\beta}$.
(a) the neutrino and (b) an example of sneutrino contribution to the
$0\nu\beta\beta$-decay amplitude $R_{0\nu\beta\beta}$.
$0\nu\beta\beta$-vertex contribution to (c) the neutrino Majorana mass
and (d) to the "Majorana"-like sneutrino mass;
(e) neutrino contribution to the sneutrino "Majorana"-like mass
and (f) sneutrino contribution to the neutrino Majorana mass.
Crossed (s)neutrino lines correspond to the (B-L)-violating propagators.

\bigskip
\noindent
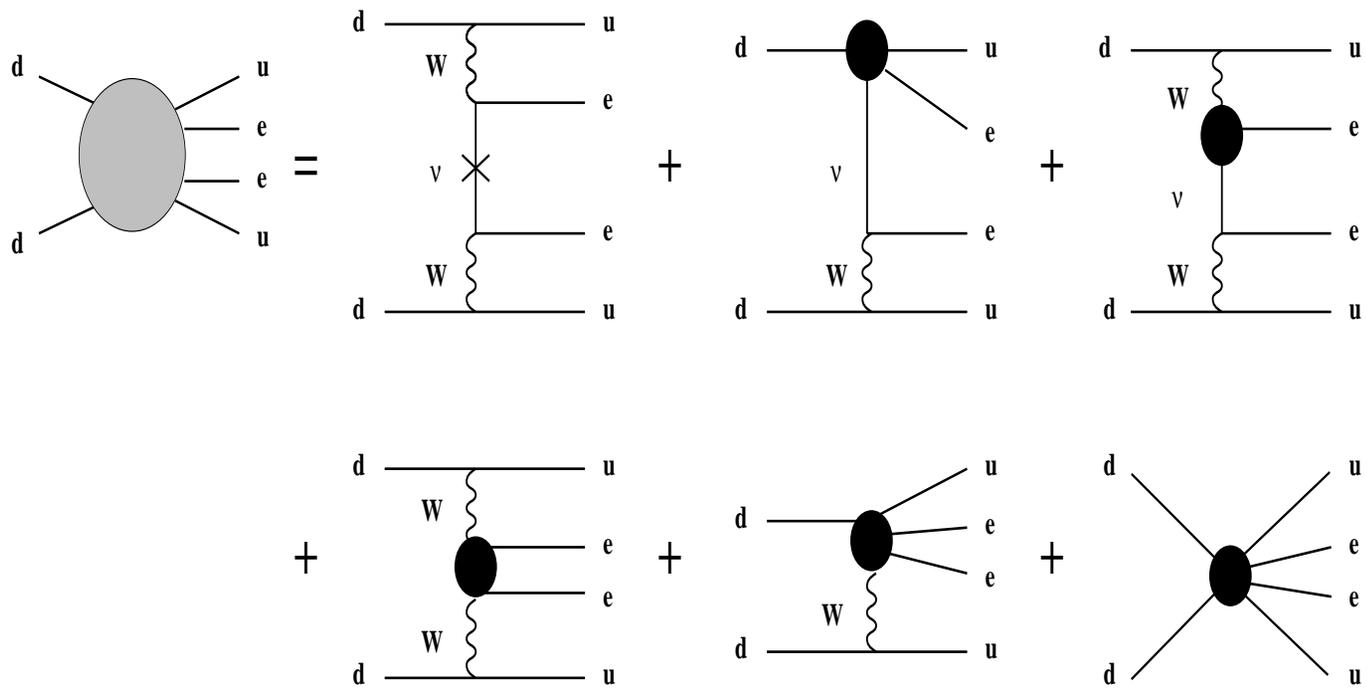
{\bf Fig.3.:} The decomposition of the $\znbb$-effective vertex.
(a) to the left: The $\znbb$-effective vertex. To the right the
decomposition: (b-d) first line: neutrino-accompanied contributions
to $\znbb$ decay, (e-g) second line: purely supersymmetric
contributions without neutrinos.

\bigskip
\noindent
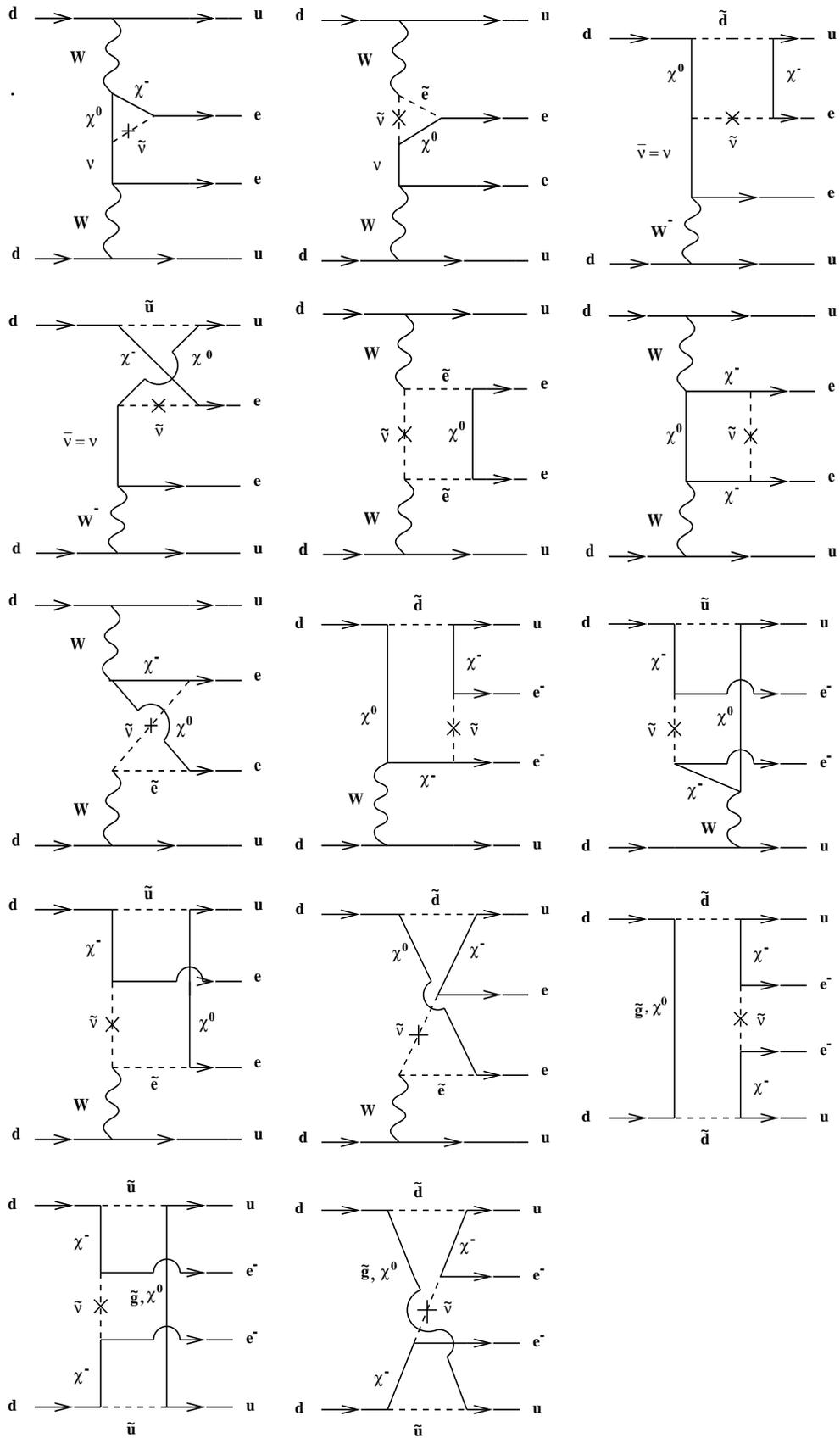
{\bf Fig.4.:} $R_P$ conserving MSSM contributions to the $\znbb$-decay
amplitude.
Leading order diagrams, see also the text.


\begin{thebibliography}{99}
%
 \bibitem{hax84} W.C. Haxton and G.J. Stephenson, Progr. Part.
                Nucl. Phys. {\bf 12}, 409 (1984);
                J. D. Vergados, Phys. Report, {\bf 133}, 1 (1986);
        K. Grotz and H.V. Klapdor-Kleingrothaus, {\it The Weak Interactions
        in Nuclear, Particle and Astrophysics}
        (Adam Hilger, Bristol, New York, 1990);
        R.N. Mohapatra and P.B. Pal, {\it Massive Neutrinos in Physics
                 and Astrophysics} (World Scientific, Singapore, 1991).
%
\bibitem{doi85} M. Doi, T. Kotani and E. Takasugi, Progr. Theor. Phys.
                        Suppl. {\bf 83}, 1  (1985).
%
\bibitem{hdmo97} Heidelberg-Moscow collaboration, M. G\"unther et al.,
                 Phys.Rev. {\bf D55} (1997) 54; and
                 J. Hellmig et al.,
                 in: {\it Proc. Int. Workshop on Dark Matter in Astro
                 and Particle Physics}, Heidelberg, Sept. 16-20, 1997,
                 World Scientific, in press

%
\bibitem{R_P-explicit}  S. Dimopoulos and L.J. Hall, Phys.Lett. {\bf B207}
                         (1987) 210;
              L. Hall and M. Suzuki, Nucl.Phys. {\bf B 231} (1984) 419;
                 E. Ma and D. Ng, Phys. Rev. {\bf D41} (1990) 1005.

\bibitem{R_P-spon}  C. Aulakh and R. Mohapatra, Phys.Lett. {\bf B119}
                    (1983) 136;
    G.G. Ross and J.W.F. Valle, Phys.Lett. {\bf B151} (1985) 375;
    J. Ellis et al., Phys.Lett. {\bf B150} (1985) 142;
    A. Santamaria and J.W.F. Valle, Phys.Lett. {\bf B195} (1987) 423;
    Phys.Rev.Lett. {\bf 60} (1988) 397; Phys.Rev. {\bf D39} (1987) 1780;
    M.C. Gonzalez-Garsia and J.W.F. Valle, Nucl.Phys. {\bf B355} (1991) 330;
    J.W.F. Valle, Phys.Lett. {\bf B196} (1987) 157;
    A. Masiero and J.W.F. Valle Phys.Lett. {\bf B251} (1990) 273.
%
\bibitem{MV} R. Mohapatra, Phys.Rev. D {\bf 34}, 3457 (1986).
%
                J.D Vergados, Phys.Lett. B {\bf 184}, 55 (1987).

\bibitem{HKK_top} M. Hirsch, H.V. Klapdor-Kleingrothaus and  S.G. Kovalenko,
                    Phys. Rev. Lett. {\bf 75}, 17 (1995).
%
\bibitem{HKK} M. Hirsch, H.V. Klapdor-Kleingrothaus and  S.G. Kovalenko,
              Phys.Lett. B {\bf 352}, 1 (1995);
              Phys. Rev. D {\bf 53}, 1329 (1996).
%
\bibitem{MB} K.S. Babu and R.N. Mohapatra,
              Phys.Rev.Lett., {\bf 75}, 2276 (1995).
%
\bibitem{HKK1} M. Hirsch, H.V. Klapdor-Kleingrothaus and S.G. Kovalenko,
               Phys.Lett. B {\bf 372}, 181 (1996);
               Erratum, Phys.Lett. B {\bf 381} (1996) 488.
\bibitem{FKSS} A. Faessler,  S.Kovalenko, F. Simkovic and J. Schwieger,
         Phys.Rev.Lett. {\bf 78} (1996) 183.
%
%
\bibitem{theorem}  M. Hirsch, H.V. Klapdor-Kleingrothaus and
                   S.G. Kovalenko, Phys. Lett. {\bf B398} (1997) 311 and
                   hep-ph/9701273
%
\bibitem{Haber} H.E. Haber and G.L.Kane, Phys.Rep. {\bf 117},  75  (1985);
                H.P. Nilles, Phys.Report. {\bf 110},  1 (1984).
%
\bibitem{SV} J. Schechter and J.W.F. Valle, Phys.Rev. D {\bf 25}, 2951 (1982);
%
            J.F. Nieves, Phys.Lett. B {\bf 147}, 375 (1984);
            E. Takasugi, Phys.Lett. B {\bf 149}, 372 (1984);
            B. Kayser, in Proc. of {\it the XXIII Int. Conf on High
            Energy Physics,} ed. S. Loken (World Scientific
            Singapore, 1987), p. 945;
            S. Petcov, in Proc. of {\it 86' Massive Neutrinos in
            Astrophysics and in Particle Physics,} ed. O. Fackler and
            J. Tr\^an Than V\^an (Editions Frontieres,
            Gif-sur-Yvette, France, 1986), p. 187;
            S.P. Rosen, UTAPHY-HEP-4 and hep-ph/9210202.

%
%
\bibitem{NRL} K. Muto and H.V. Klapdor, in: {\it Neutrinos},
              Springer, Heidelberg, 1988, p.183

%
\bibitem{Vergados2} J.D. Vergados, Phys.Rev.{\bf C24} (1981) 640;
                                   Nucl.Phys.{\bf B 218} (183) 109.
%
\bibitem{lrmodel} M. Hirsch, H.V. Klapdor-Kleingrothaus and O. Panella,
                  Phys.Lett. {\bf B 374} (1996) 7;
                  M. Hirsch and H.V. Klapdor-Kleingrothaus,
                  Proc. Int. Workshop on {\it Double Beta
                  Decay and related topics}, Trento, Italy,
                  April 24. - May 5., 1995, eds.
                  H.V. Klapdor-Kleingrothaus and S. Stoica, World
                  Scientific, 1996, p. 175
%
\bibitem{mut89} K. Muto, E. Bender and H.V. Klapdor, Z. Phys. {\bf A 334}
                (1989) 177; {\it ibid} 187;
                M. Hirsch, K. Muto, T. Oda and H.V. Klapdor-Kleingrothaus,
                Z. Phys. {\bf A 347} (1994) 151
%
\bibitem{Jungman} For recent review see for instance:
                G. Jungman, M. Kamionkowski and K. Griest,
                  Phys.Rept. {\bf 267} (1996) 195.
%
\bibitem{PDG} Particle Data Group, Phys.Rev. {\bf D54}  (1996) 1-720.
%

\bibitem{sn_accel} M. Hirsch, H.V. Klapdor-Kleingrothaus, St. Kolb and
                    S.G. Kovalenko, submitted to Phys.Rev. {\bf D}
%
\bibitem{Grossman_Haber} Y. Grossman and H.E. Haber, 
                         Phys. Rev. Lett. {\bf 78} (1997) 3438 

%
\bibitem{gun86} J.F. Gunion, H.E. Haber, G.L. Kane,
                Nucl.Phys. {\bf B 272} (1986) 1.
%
\end{thebibliography}
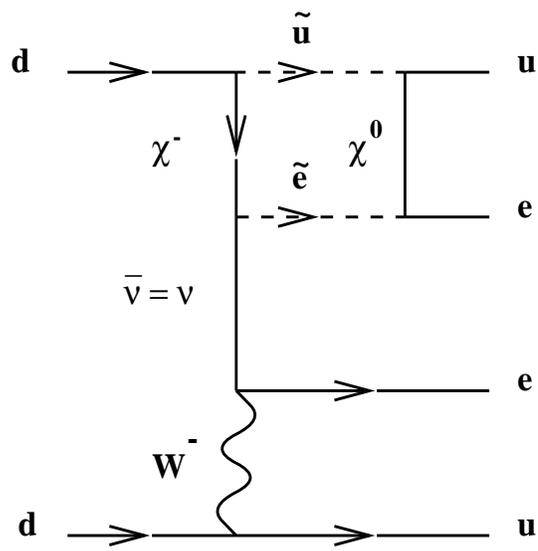
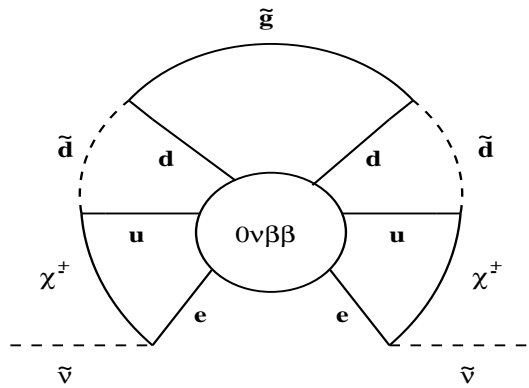
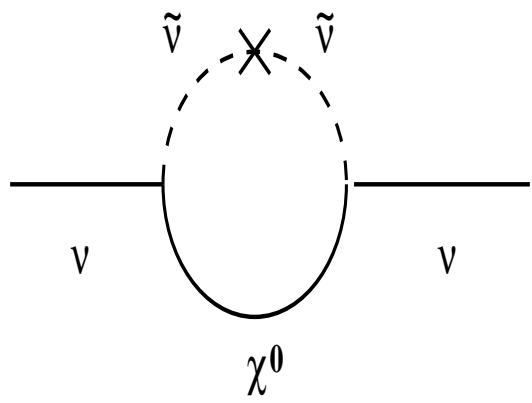
\end{document}